\documentclass[%
reprint,
superscriptaddress,
amsmath,amssymb,
aps,
prl,
]{revtex4-2}

\usepackage{times}

\usepackage{graphicx}
\usepackage{dcolumn}
\usepackage{bm}
\usepackage[utf8]{inputenc}

\usepackage{easyReview}

\graphicspath{{img/}}

\begin{document}
	
\preprint{APS/123-QED}

\title{Porosity and Material Disorder Drive Distinct Channelization Transition}

\author{Andr\'{e} F.~V. Matias}
\email{a.f.valentematias@uu.nl}
\affiliation{Centro de F\'{i}sica Te\'{o}rica e Computacional, Faculdade de Ci\^{e}ncias, Universidade de Lisboa, 1749--016 Lisboa, Portugal}
\affiliation{Soft Condensed Matter \& Biophysics, Debye Institute for Nanomaterials Science, Utrecht University, Princetonplein 1, 3584 CC Utrecht, The Netherlands}
\author{Rodrigo C.~V. Coelho}
\affiliation{Centro de F\'{i}sica Te\'{o}rica e Computacional, Faculdade de Ci\^{e}ncias, Universidade de Lisboa, 1749--016 Lisboa, Portugal}
\affiliation{Centro Brasileiro de Pesquisas Físicas, Rua Xavier Sigaud 150, 22290-180 Rio de Janeiro, Brazil}
\author{Humberto A. Carmona}
\affiliation{Departamento de F\'{i}sica, Universidade Federal do Cear\'{a}, 60451--970, Fortaleza, Cear\'{a}, Brazil}
\author{Jos\'{e} S. Andrade Jr.}
\affiliation{Departamento de F\'{i}sica, Universidade Federal do Cear\'{a}, 60451--970, Fortaleza, Cear\'{a}, Brazil}
\author{Nuno A.~M. Ara\'{u}jo}
\email{nmaraujo@fc.ul.pt}
\affiliation{Centro de F\'{i}sica Te\'{o}rica e Computacional, Faculdade de Ci\^{e}ncias, Universidade de Lisboa, 1749--016 Lisboa, Portugal}
\affiliation{Departamento de F\'{i}sica, Faculdade de Ci\^{e}ncias, Universidade de Lisboa, 1749--016 Lisboa, Portugal}


\begin{abstract}
    
Flow through porous media can reshape the medium through erosion and deposition, producing preferential flow channels across a wide range of natural and industrial systems. Yet the mechanisms by which spatial disorder triggers channelization remain unclear. Here we derive a continuum description for the coupled evolution of flow and porosity by coarse-graining pore-scale dynamics and validating the resulting model with pore-scale simulations. Using this framework, we show that different sources of disorder lead to qualitatively distinct behaviors. Disorder in erosion resistance produces a discontinuous transition to localized flow, with permanent channels appearing only above a finite disorder strength. In contrast, even extremely weak fluctuations in the initial porosity destabilize homogeneous flow and trigger persistent channelization. These results reveal an unexpected sensitivity of evolving porous media to structural heterogeneity, suggesting that channelization can arise generically even in nearly uniform materials.
\end{abstract}

\maketitle

Water flowing downhill changes the soil through erosion, dissolution, and deposition, leading to intricate networks of channels at different scales, such as braided or continental river networks~\cite{Banavar1997, Derr2020, Hilton2020}. The same happens under the surface. Soils are heterogeneous mixtures of different materials forming porous structures through which water penetrates, changing the medium and forming channels~\cite{Mahadevan2012, Ferlito2006, Moore2023, Xu2018, Cerasi1998, Tauber2026}. The size and shape of these channels vary significantly with the soil properties and range from nanochannels in shale rocks~\cite{Sun2020, Zhu2019, Zwietering1954} to large aquifers~\cite{Lefebvre2016, Maliva2015, D4SM00391H}. Understanding the dynamics of flow-induced channelization in porous media is of paramount relevance, not only in geophysics but also for several industrial processes, such as oil extraction~\cite{Ghassemi2015, Ladd2021, Rassamdana1996}, carbon capture~\cite{Schlesinger1999, Oost2007, Borrelli2017, Lee2025}, filtering~\cite{Mondal2019, Dalwadi2016, Araujo2006}, and food processing~\cite{Mo2021}.

\begin{figure}[!ht]
	\centering
	\includegraphics[width=\linewidth]{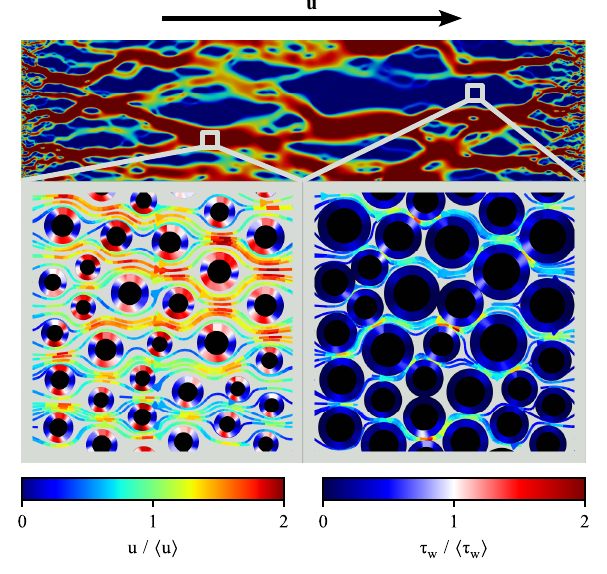}
	\caption[Example of velocity and porosity fields coupling]{\textbf{Example of velocity and porosity fields coupling.} (Top) Example of a velocity field obtained with the continuum model for the coupled velocity and porosity fields, starting from a weak hyperbolic disorder in porosity. The colors correspond to the magnitude of the velocity (left scale). Mild initial differences are amplified since the stronger shear stress in regions of large porosity triggers erosion, while the low stresses in small pores favor deposition. (Bottom) Examples of two streamlines and distribution of the wall shear stresses obtained through pore-scale simulations of a system consisting of randomly distributed circular obstacles (black circles). The colors represent the magnitude of the velocity in the fluid regions (left scale) and the wall shear stress on the surface of the obstacles (right scale). These are examples of regions as those highlighted in the velocity field on top.}
	\label{fig:pore_to_field}
\end{figure}

\begin{figure*}
	\centering
	\includegraphics[width=1\linewidth]{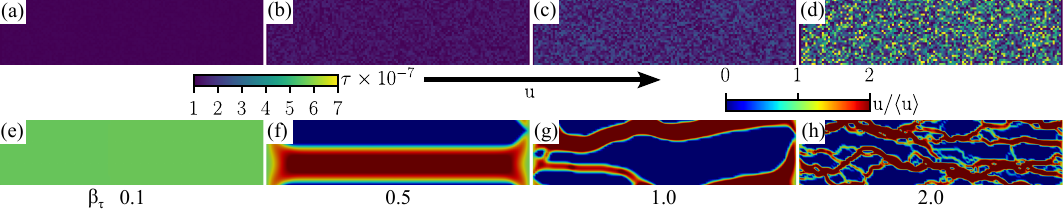}
	\caption[Erosion resistance and steady-state velocity fields for different strengths of hyperbolic disorder on a lattice with $128\times32$ nodes]{\textbf{Erosion resistance and steady-state velocity fields for different strengths of hyperbolic disorder on a lattice with $128\times32$ nodes.} (a)-(d) Erosion resistance field, where the color represents $\tau$ (left scale). Weak disorder corresponds to $\beta_\tau=0.1$ and $0.5$, while strong disorder corresponds to $\beta_\tau=2$, generating a broader range of erosion resistance values. (e)-(h) Steady-state velocity field for the same disorder strengths. Colors represent the velocity normalized by the average velocity (right scale), where the average velocity corresponds to the imposed inlet and outlet velocity. For weak disorder the flow is nearly homogeneous, whereas for strong disorder the flow localizes in narrow channels with velocities much larger than the average.}
	\label{fig:vel_field}
\end{figure*}

Experimental and numerical studies have shown that channelization emerges from a truly multiscale dynamics, with local changes in the pore structure triggering macroscopic flow redistributions~\cite{Kudrolli2016, Mahadevan2012, Zareei2021, Ocko2015, Maionchi2008, Zhang2023}. A detailed model at the pore-scale will fail to access the relevant length and time scales of the redistribution~\cite{Menke2023, Jager2017}. Here, instead, we coarse grain the local dynamics and derive a continuum model for the coupled velocity and porosity fields. Erosion and deposition are driven by fluid-induced shear stresses on the solid-liquid interface~\cite{Parker2000, Mo2021}, which is well captured by evolving capillaries, as shown by pore-scale simulations~\cite{Matias2021}. To access larger scales while reducing the computational effort, we calculate the shear stresses from the (local) Reynolds number. This enables us to study how spatial disorder controls the onset of channelization. The spatial disorder can manifest itself as variations in porosity~\cite{Sahimi2011}, as in compacted spheres~\cite{Cerasi1998} or food powders~\cite{Illy2005}, or in erosion resistance, as in soils~\cite{Sahimi1993, Bear1988}. While heterogeneity in material properties is often assumed to be the primary mechanism driving channel formation, we show that the dynamics is far more sensitive to disorder in porosity. Using a continuum model validated with pore-scale simulations, we demonstrate that even extremely weak porosity fluctuations can destabilize homogeneous flow and trigger channelization, whereas disorder in erosion resistance requires a finite critical strength (see Fig.~\ref{fig:pore_to_field}).

Fluid flow through a porous medium is described by the generalized Navier-Stokes equations~\cite{Nithiarasu1997}:
\begin{subequations}
\label{eq:gen_navier_stokes}
\begin{equation}
	\boldsymbol{\nabla} \cdot \mathbf{u}=0,
\end{equation}
\begin{equation}
	\frac{\partial \mathbf{u}}{\partial t}+(\mathbf{u} \cdot \boldsymbol{\nabla})\left(\frac{\mathbf{u}}{\phi}\right)=-\frac{1}{\rho} \boldsymbol{\nabla}(\phi p)+\nu \nabla^2 \mathbf{u}+\mathbf{F} \\,
\end{equation}
\end{subequations}
where $\mathbf{u}$, $p$, and $\phi$ are the Darcy-scale velocity, pressure, and porosity fields, respectively, $\rho$ is the fluid density, and $\nu$ is the kinematic viscosity. $\mathbf{F}$ is the net body force, given by,
\begin{equation}
	\mathbf{F}=-\frac{\phi \nu}{k} \mathbf{u} + \phi \mathbf{G},
\end{equation}
where $k$ is the permeability of the medium, $\mathbf{G}$ is the external force field, and we have neglected higher-order terms in the velocity field. The permeability depends on the pore structure~\cite{Graczyk2020}, which we describe with the capillary model, assuming a Poiseuille flow across capillaries of radius $a$~\cite{Feder2022}. For packed spheres, the permeability is commonly approximated by the semi-empirical Kozeny–Carman relation~\cite{Kozeny1927, Carman1997}
\begin{equation}
	k=\frac{a^2}{18} \frac{\phi^3}{(1-\phi)^2} \\,
	\label{eq:permability}
\end{equation}
where $a$ and $\phi$ are functions of position and time.

The rates of erosion and deposition are linear functions of the shear stress at the wall $\tau_w=\mu(\partial v / \partial y)|_\mathrm{wall}$, where $v$ is the component of the pore-scale velocity parallel to the solid surface and $y$ is the closest distance to wall~\cite{Acheson1990,Jager2017, PhysRevFluids.9.114301}. Thus,
\begin{equation}
	\dot{a} =
	\begin{cases}
		-\kappa_{\mathrm{dep}}\left(\tau_{\mathrm{dep}}-\tau_w\right) & \tau_w < \tau_{\mathrm{dep}} , \\
        0 & \tau_{\mathrm{dep}} \le \tau_w \le \tau_{\mathrm{er}} , \\
        +\kappa_{\mathrm{er}}\left(\tau_w-\tau_{\mathrm{er}}\right) & \tau_{\mathrm{er}} < \tau_w ,
	\end{cases}
	\label{eq:cap_evolution}
\end{equation}
where $\kappa_{\mathrm{dep}}$ and $\kappa_{\mathrm{er}}$ are the density dependent rates that set the time scales of the deposition and erosion, respectively, and $\tau_{\mathrm{dep}}$ and $\tau_{\mathrm{er}}$ are the corresponding thresholds. This relates to the porosity as $\dot{\phi}/\phi = \dot{a}/a$. Hereafter, we assume that \mbox{$\kappa_{\mathrm{dep}} = \kappa_{\mathrm{er}} = \kappa$} and $\tau_{\mathrm{dep}} = \tau_{\mathrm{er}} = \tau$. Clogging by fine powders is neglected and the fluid is assumed to be saturated, so there is always solute for deposition. Assuming parallel plates, we obtain
\begin{equation}
	\tau_w = 6 \rho \frac{\| \mathbf{u}\|^2}{\textrm{Re}} \\,
	\label{eq:wss}
\end{equation}
where $\textrm{Re} = (\| \mathbf{u}\| \langle h\rangle)/\nu$ is the (local) Reynolds number, $\langle h\rangle$ is the average pore size, which we assume to be equal to $a$, and $\| \mathbf{u}\|$ is the Darcy velocity. The predicted relation is validated by pore-scale simulations for 2D random arrangements of circular obstacles (see End Matter), and a more detailed study can be found in Ref.~\cite{Kahza2024}.

\begin{figure*}
	\centering
	\includegraphics[width=\linewidth]{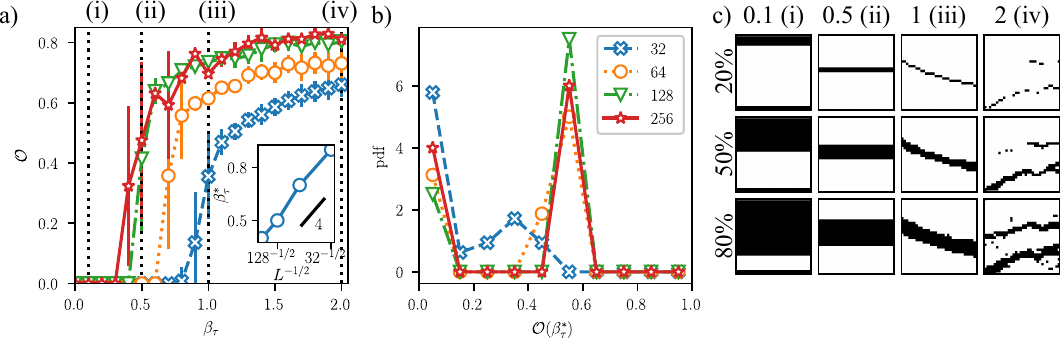}
	\caption[Disorder on the erosion resistance]{\textbf{Disorder on the erosion resistance.} (a) Channelization parameter $\mathcal{O}$ as a function of the strength of disorder in the erosion resistance, for different system sizes (colors and markers). Each point is an average over $N$ steady-state samples, decreasing systematically with the system size down to a minimum of $N=4$. Error bars correspond to the standard error. For weak disorder, the channelization parameter is zero, as the fluid flow is homogeneous, see Fig.~\ref{fig:vel_field}(e). Increasing the strength of disorder increases the channelization parameter, as small channels of larger velocity emerge, see Fig.~\ref{fig:vel_field}(h). The inset contains the log-scale plot of the strength of disorder that corresponds to the maximum second momentum of $\mathcal{O}$ as a function of the inverse square root of the system size $L^{-1/2}$. (b) Steady-state probability density distribution (pdf) of the channelization parameter for the transition disorder strength ($\beta_\tau=\beta^*_\tau$) for different system sizes (colors and markers). The pdf is characterized by a bimodal distribution characteristic of a system that exhibits a first-order phase transition. (c) In black, the central nodes ($x\in[3/8L_x,5/8L_x]$) that accumulate $20\%$, $50\%$, and $80\%$ of the flow rate (rows) for $\beta_\tau=\{0.1, 0.5, 1, 2\}$ (columns and vertical lines in (a)) for a system with length $128$ nodes. For weak disorder strength, the flow is distributed across a single wide channel (i). At the transition disorder strength, the channel is close to its minimum size (ii). Increasing the disorder strength past the transition point, the channel becomes increasingly tortuous in order to find the optimal path that maximizes hydraulic conductivity (iii). Further increasing the disorder strength, the flow splits into multiple small channels separated by regions with very high resistance to erosion.}
    \label{fig:threshold_disorder}
\end{figure*}

Let us first consider a system that has constant initial porosity ($\phi_0=0.5$) but is composed of a mixture of materials with different erosion resistance $\tau$. We assume constant $\kappa=10^{-7}$ lattice units (l.u), and resolve the time evolution of the velocity, pressure, and porosity fields, by integrating Eqs.~\eqref{eq:gen_navier_stokes}-\eqref{eq:cap_evolution} on a discrete lattice using the lattice Boltzmann method for the fluid flow and Euler method for the porosity time evolution~\cite{Guo2002a, Kruger2017, Mei1999}. The fluid flow is driven by imposing constant inlet and outlet velocity $\langle\mathbf{u}\rangle=10^{-6}$ l.u. We present all values using lattice units that can be converted into real units by comparing with dimensionless numbers like the Reynolds or Péclet number. The initial erosion resistance in each position is generated at random from a hyperbolic distribution,
\begin{equation}
	p(\tau_i)\propto 1/\tau_i \\,
    \label{eq:power_law}
\end{equation}
truncated between $\tau_{\min}=10^{-7}$ l.u. and $\tau_{\max}/\tau_{\min}=\exp(\beta_\tau)$, where $\beta_\tau \ge 0$ is the strength of disorder, with $\tau_{\max}\rightarrow\tau_{\min}$, when $\beta_\tau \rightarrow 0 $ (see further details in the End Matter and Ref.~\cite{Andrade2009}). Figures~\ref{fig:vel_field}(a)-(d) show the erosion resistance field for different values of disorder $\beta_\tau$. For low values of $\beta_\tau$ (weak disorder), the truncated distribution is very narrow and all values of erosion resistance are close to $\tau_{\min}$, as shown in Fig.~\ref{fig:vel_field}(a). As $\beta_\tau$ increases, the distribution of the initial erosion resistance becomes broader, which corresponds to stronger disorder (see Figs.~\ref{fig:vel_field}(c)-(d)). The erosion resistance distribution is uncorrelated in space. However, in regions of lower erosion resistance, the porosity shall increase. By contrast, in regions of higher erosion resistance, the porosity decreases through deposition (see also Fig.~\ref{fig:pore_to_field}). This evolution of the porosity is then reflected on the velocity field, in such a way that regions with higher porosity correspond to regions of higher velocity and consequently higher shear stress. Thus, small differences in the erosion resistance are amplified in the velocity and porosity fields, creating correlations and a dynamic flow redistribution until a steady-state is reached, where the porosity distribution is bimodal and spatial correlations emerge (see Supplementary Fig.~{S1}-{S2}~\cite{SM}). The velocity field in the steady-state for the different values of $\beta_\tau$ is shown in Figs.~\ref{fig:vel_field}(e)-(h). For the lowest values, wide channels are formed, since erosion occurs almost everywhere, as shown in Fig.~\ref{fig:vel_field}(e)-(f). As the erosion resistance distribution gets wider (stronger disorder), there are regions in which deposition takes place and, over time, the fluid flow is diverted to regions of higher porosity. This increases the tortuosity of the main channel as it seeks for the optimal path that maximizes hydraulic conductivity (Fig.~\ref{fig:vel_field}(g)). When the disorder is sufficiently strong, low-porosity regions emerge that divide the flow into separate channels (see Fig.~\ref{fig:vel_field}(h) and Supplementary Fig.~{S2}(f)~\cite{SM}).

To quantify the spatial distribution of the fluid velocity, we introduce a channelization parameter $\mathcal{O}$, defined as,
\begin{equation}
	\mathcal{O} = 1 - \frac{\langle e\rangle^2}{\left\langle e^2\right\rangle} \\,
	\label{eq:order_param}
\end{equation}
where $\left\langle e^n\right\rangle=(1 / V) \iint(\mathbf{u} \cdot \mathbf{u})^n d^2 \mathbf{r}$ is the $n^\textrm{th}$ moment of the kinetic energy, and $V$ is the number of lattice nodes in the integration region. Note that, the second term is the usual participation ratio~\cite{Andrade1999, Soares1999, Seybold2020, Araujo2013}. If the kinetic energy is homogeneous in space, $\mathcal{O} \rightarrow 0$, while for strongly localized flows we expect that $\mathcal{O} \rightarrow 1-1/V$, and thus converges to one in the thermodynamic limit. As shown in Fig.~\ref{fig:threshold_disorder}(a), $\mathcal{O}$ is a monotonic increasing function of the strength of disorder $\beta_\tau$. This is consistently observed for different system sizes. For weak disorder the flow is homogeneous in space and for strong disorder the flow is localized in small channels. In between, there is a transition that sharpens with increasing system size, showing that there is a threshold value of disorder below which no channelization occurs. This sharp increase of $\mathcal{O}$ with $\beta_\tau$ signals the onset of flow localization, suggesting a transition for which $\mathcal{O}$ serves as an order parameter. In the inset of Fig.~\ref{fig:threshold_disorder}(a) we show that the disorder value of the transition $\beta^*_\tau$, estimated as the value of $\beta_\tau$ that maximizes $\langle\mathcal{O}^2(\beta_\tau)\rangle - \langle\mathcal{O}(\beta_\tau)\rangle^2$, exhibits an approximately linear dependence on $L^{-1/2}$ over the range of system sizes considered. A linear extrapolation yields $\beta^*_\tau(L\rightarrow\infty)=0.12$. This, together with the bimodal distribution of the probability density function of $\mathcal{O}$ at the transition, as shown in Fig.~\ref{fig:threshold_disorder}(b), are consistent with a discontinuous transition, although additional finite-size analysis would be required to establish this more firmly. The change in the flow dynamics before and after the transition is shown in the snapshots of Fig.~\ref{fig:threshold_disorder}(c) that contain, in black, the nodes for each vertical line carrying $20\%$, $50\%$, and $80\%$ of the total flux at the steady-state, and for distinct values of $\beta$ in the horizontal lines, namely $\beta=0.1$ (i), $0.5$ (ii), $1.0$ (iii) and $2.0$ (iv). While for low values of disorder (i) the flow is uniformly distributed in space, for $\beta_\tau=2$ (iv), the flow is localized in a few tortuous channels.

\begin{figure}
	\centering
	\includegraphics[width=\linewidth]{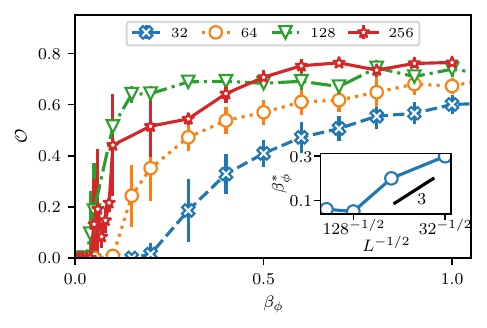}
    \caption[Disorder on the initial porosity]{\textbf{Disorder on the initial porosity.} Channelization parameter $\mathcal{O}$ as a function of the strength of disorder in the initial porosity, for different system sizes (colors and markers). Each point is an average over $N$ steady-state samples, with $N$ decreasing with increasing system size down to a minimum of $N=4$. Error bars correspond to the standard error.    The inset contains the log-scale plot of the strength of disorder that corresponds to the maximum second momentum of $\mathcal{O}$ as a function of the inverse square root of the system size $L^{-1/2}$.}
	\label{fig:porosity_disorder}
\end{figure}

We now consider disorder in the initial porosity. We integrate Eq.~\eqref{eq:gen_navier_stokes}-\eqref{eq:wss} with a uniform erosion resistance $\tau=10^{-7}$~l.u., but with a truncated hyperbolic distribution of the initial porosity $\phi_0$. The distribution is described by \mbox{$\phi_0/\phi_{\max}\in]\exp(-\beta_\phi), 1]$}, where $\beta_\phi$ determines the disorder strength and $\phi_\textrm{max}=0.5$ sets the maximum initial porosity. As in the previous case, the degree of channelization depends on the disorder strength, with the value at which localization emerges depending on the system size, as shown in the inset of Fig.~\ref{fig:porosity_disorder}, although with a lower slope compared to the case of disorder in erosion resistance. For the system sizes investigated, the estimated transition point becomes weakly dependent on system size for $L\ge128$, suggesting an asymptotic value $\beta^*_\phi(L\rightarrow\infty)\approx0.05$. A similar trend is observed for systems with lower erosion resistance, for which $\beta^*_\phi(L\rightarrow\infty)\approx0.01$, as shown in Supplementary Fig.~{S3}.

Because permeability depends nonlinearly on porosity, even weak heterogeneities are strongly amplified by the coupled erosion–flow feedback, leading to an effective destabilization of homogeneous flow. As a consequence, finite-size effects are weaker and the scaling behavior is less pronounced than in the case of erosion disorder, consistent with a near-marginal instability of the uniform state. While disorder in erosion resistance displays signatures compatible with a discontinuous localization transition, porosity disorder leads instead to a much softer onset of channelization.

In conclusion, we studied how different sources of disorder trigger channelization in evolving porous media subject to erosion and deposition. We considered systems with disorder in erosion resistance, as in heterogeneous mixtures such as soils, cement, and sedimentary rocks~\cite{Bear1988, Goretta1999, Birien2023}. As intuitively expected, the fluid–structure interaction results in flow localization: erosion occurs preferentially in regions of low erosion resistance, while deposition occurs in regions of high erosion resistance. Nevertheless, flow localization requires a finite critical disorder strength, leading to flow confined in localized channels.

Even for uniform erosion resistance, channelization can emerge from spatial imbalance in the initial porosity, as reported in previous pore-scale works~\cite{Cerasi1998,Moore2023,Jager2017}. Here we show that this mechanism is also present at the field-scale, and that the dynamics is considerably more sensitive to porosity heterogeneity than to disorder in erosion resistance. In particular, channelization induced by porosity disorder is triggered by extremely weak fluctuations, with the initial porosity such that $\phi_0/\phi_{\max}\in[0.99,1]$. Within accessible system sizes, the apparent threshold is very small and only weakly size-dependent, indicating that homogeneous flow is marginally stable with respect to porosity fluctuations. As such, channelization is expected to be widespread in natural porous media, even in systems composed of uniform materials.

This work is made possible by a continuum model that incorporates erosion and deposition at the Darcy-scale, allowing for field-scale simulations and enabling access to length and time scales compatible with geological, industrial, and filtration applications. The model assumes the Kozeny-Carman approximation for the permeability of the evolving porous system to determine the wall shear stress, and was validated with pore-scale simulations of random packings of spheres undergoing erosion and deposition, as described in the End Matter. All simulations, both pore-scale and field-scale, were conducted in two-dimensional systems and performed using the lattice Boltzmann method, but we expect a straightforward implementation of our shear-stress model in other methods, such as Finite Element and Finite Volume, and extension to three dimensions, as discussed in the End Matter, where no qualitative differences are expected. Furthermore, our model allows the investigation of porous media heterogeneity beyond the hyperbolic disorder considered here. The proposed model neglects the shape and size of eroded solid particles, which are typically fine and transported by the flow. When particle sizes become comparable to the characteristic pore length, clogging of narrow channels is expected. Although this mechanism is not included in the present model, it is likely to further enhance localization by reducing the local hydraulic conductivity of low-porosity regions, while leaving larger pores largely unaffected.

\begin{acknowledgments}
    
We acknowledge financial support from the Portuguese Foundation for Science and Technology (FCT) under the contracts no. UIDB/00618/2020 (DOI 10.54499/UIDB/00618/2020), UIDP/00618/2020 (DOI 10.54499/UIDP/00618/2020), SFRH/BD/143955/2019, 2023.10412.CPCA.A2 (DOI 10.54499/2023.10412.CPCA.A2), FCT/Mobility/1348751812/2024-25, and UID/00618/2025 (DOI 10.54499/UID/00618/2025). AFVM acknowledges funding from the European Union’s Horizon Europe research and innovation program under the grant agreement number 101203506, Marie
Sklodowska-Curie Action Postdoctoral Fellowship,
project IonFlowElast. RC acknowledges the financial support from FAPERJ – Fundação Carlos Chagas Filho de Amparo à Pesquisa do Estado do Rio de Janeiro (Processo SEI-260003/020878/2025). HAC and JSA acknowledge the Brazilian agencies CNPq, CAPES, and FUNCAP for financial support.
\end{acknowledgments}

\nocite{*}
\bibliography{ref}

@article{Guo2002a,
   author = {Zhaoli Guo and T. S. Zhao},
   doi = {10.1103/PhysRevE.66.036304},
   issue = {3},
   journal = {Physical Review E},
   pages = {036304},
   publisher = {American Physical Society},
   title = {Lattice Boltzmann model for incompressible flows through porous media},
   volume = {66},
   year = {2002},
}

@article{PhysRevFluids.9.114301,
  title = {Mathematical modeling of deposition and erosion in porous media with branching channels},
  author = {Mazi, Emeka Peter and El Kahza, Hamad and Sanaei, Pejman},
  journal = {Phys. Rev. Fluids},
  volume = {9},
  issue = {11},
  pages = {114301},
  numpages = {28},
  year = {2024},
  month = {Nov},
  publisher = {American Physical Society},
  doi = {10.1103/PhysRevFluids.9.114301},
  url = {https://link.aps.org/doi/10.1103/PhysRevFluids.9.114301}
}

@Article{D4SM00391H,
author ="Voigtländer, Anne and Houssais, Morgane and Bacik, Karol A. and Bourg, Ian C. and Burton, Justin C. and Daniels, Karen E. and Datta, Sujit S. and Del Gado, Emanuela and Deshpande, Nakul S. and Devauchelle, Olivier and Ferdowsi, Behrooz and Glade, Rachel and Goehring, Lucas and Hewitt, Ian J. and Jerolmack, Douglas and Juanes, Ruben and Kudrolli, Arshad and Lai, Ching-Yao and Li, Wei and Masteller, Claire and Nissanka, Kavinda and Rubin, Allan M. and Stone, Howard A. and Suckale, Jenny and Vriend, Nathalie M. and Wettlaufer, John S. and Yang, Judy Q.",
title  ="Soft matter physics of the ground beneath our feet",
journal  ="Soft Matter",
year  ="2024",
volume  ="20",
issue  ="30",
pages  ="5859-5888",
publisher  ="The Royal Society of Chemistry",
doi  ="10.1039/D4SM00391H"
}

@article{Mei1999,
   author = {Renwei Mei and Li-Shi Luo and Wei Shyy},
   doi = {10.1006/JCPH.1999.6334},
   issn = {0021-9991},
   issue = {2},
   journal = {Journal of Computational Physics},
   pages = {307-330},
   publisher = {Academic Press},
   title = {An Accurate Curved Boundary Treatment in the Lattice Boltzmann Method},
   volume = {155},
   year = {1999},
}

@book{Kruger2017,
   author = {Timm Krüger and Halim Kusumaatmaja and Alexandr Kuzmin and Orest Shardt and Gonçalo Silva and Erlend Magnus Viggen},
   city = {Cham},
   doi = {10.1007/978-3-319-44649-3},
   edition = {1},
   isbn = {978-3-319-44649-3},
   publisher = {Springer Cham},
   title = {The Lattice Boltzmann Method, Principles and Practice},
   year = {2017},
}

@book{Feder2022,
   author = {Jens Feder and Eirik Grude Flekkøy and Alex Hansen},
   doi = {10.1017/9781009100717},
   isbn = {9781108989114},
   journal = {Physics of Flow in Porous Media},
   publisher = {Cambridge University Press},
   title = {Physics of Flow in Porous Media},
   year = {2022},
}

@article{Seybold2020,
   author = {H. J. Seybold and H. A. Carmona and F. A. {Leandro Filho} and A. D. Araújo and F. {Nepomuceno Filho} and José S. {Andrade Jr.}},
   doi = {10.1103/PhysRevFluids.5.104101},
   issue = {10},
   journal = {Physical Review Fluids},
   pages = {104101},
   publisher = {American Physical Society},
   title = {Flow through three-dimensional self-affine fractures},
   volume = {5},
   year = {2020},
}

@article{Araujo2013,
   author = {Nuno A. M. Araújo and H. Seybold and R. M. Baram and Hans J. Herrmann and José S. {Andrade Jr.}},
   doi = {10.1103/PhysRevLett.110.064106},
   issue = {6},
   journal = {Physical Review Letters},
   pages = {064106},
   publisher = {American Physical Society},
   title = {Optimal Synchronizability of Bearings},
   volume = {110},
   year = {2013},
}

@article{Nithiarasu1997,
   author = {P. Nithiarasu and T. Sundararajan and K. N. Seetharamu},
   doi = {10.1016/S0735-1933(97)00106-1},
   issue = {8},
   journal = {International Communications in Heat and Mass Transfer},
   pages = {1121-1130},
   publisher = {Pergamon},
   title = {Double-diffusive natural convection in a fluid saturated porous cavity with a freely convecting wall},
   volume = {24},
   year = {1997},
}

@article{Rabbani2014,
   author = {Arash Rabbani and Saeid Jamshidi and Saeed Salehi},
   doi = {10.1016/j.petrol.2014.08.020},
   journal = {Journal of Petroleum Science and Engineering},
   pages = {164-171},
   publisher = {Elsevier},
   title = {An automated simple algorithm for realistic pore network extraction from micro-tomography images},
   volume = {123},
   year = {2014},
}

@article{Jager2017,
   author = {Robin Jäger and M. Mendoza and Hans J. Herrmann},
   doi = {10.1103/PhysRevE.95.013110},
   issue = {1},
   journal = {Physical Review E},
   month = {1},
   pages = {013110},
   title = {Channelization in porous media driven by erosion and deposition},
   volume = {95},
   year = {2017},
}

@article{Parker2000,
   author = {Gary Parker and Norihiro Izumi},
   doi = {10.1017/S0022112000001403},
   journal = {Journal of Fluid Mechanics},
   pages = {203-238},
   publisher = {Cambridge University Press},
   title = {Purely erosional cyclic and solitary steps created by flow over a cohesive bed},
   volume = {419},
   year = {2000},
}

@article{Menke2023,
   author = {Hannah P. Menke and Julien Maes and Sebastian Geiger},
   doi = {10.1038/s41598-023-37725-6},
   issn = {2045-2322},
   issue = {1},
   journal = {Scientific Reports},
   pages = {11312},
   title = {Channeling is a distinct class of dissolution in complex porous media},
   volume = {13},
   year = {2023},
}

@article{Matias2021,
   author = {André F. V. Matias and Rodrigo C. V. Coelho and José S. {Andrade Jr.} and Nuno A. M. Araújo},
   doi = {10.1016/j.jocs.2021.101360},
   issn = {18777503},
   journal = {Journal of Computational Science},
   pages = {101360},
   title = {Flow through time–evolving porous media: Swelling and erosion},
   volume = {53},
   year = {2021},
}

@article{Zareei2021,
   author = {Ahmad Zareei and Deng Pan and Ariel Amir},
   doi = {10.1103/PhysRevLett.128.234501},
   issn = {0031-9007},
   issue = {23},
   journal = {Physical Review Letters},
   pages = {234501},
   title = {Temporal Evolution of Erosion in Pore Networks: From Homogenization to Instability},
   volume = {128},
   year = {2022},
}

@article{Ghassemi2015,
   author = {Ali Ghassemi and Ali Pak},
   doi = {10.1016/J.PETROL.2015.09.019},
   issn = {0920-4105},
   journal = {Journal of Petroleum Science and Engineering},
   pages = {218-231},
   publisher = {Elsevier},
   title = {Numerical simulation of sand production experiment using a coupled Lattice Boltzmann -- Discrete Element Method},
   volume = {135},
   year = {2015},
}

@article{Schlesinger1999,
   author = {William H. Schlesinger},
   doi = {10.1126/SCIENCE.284.5423.2095/SUPPL_FILE/1042216.XHTML},
   issn = {00368075},
   issue = {5423},
   journal = {Science},
   pages = {2095},
   publisher = {American Association for the Advancement of Science},
   title = {Carbon sequestration in soils},
   volume = {284},
   year = {1999},
}

@book{Illy2005,
   author = {Andrea Illy and Rinantonio Viani},
   isbn = {9780123703712 9},
   pages = {398},
   publisher = {Elsevier Academic},
   title = {Espresso Coffee: The Science of Quality},
   year = {2005},
}

@book{Bear1988,
   author = {Jacob Bear},
   isbn = {0486656756},
   pages = {764},
   publisher = {Dover},
   title = {Dynamics of fluids in porous media},
   year = {1988},
}

@article{Derr2020,
   author = {Nicholas J. Derr and David C. Fronk and Christoph A. Weber and Amala Mahadevan and Chris H. Rycroft and L. Mahadevan},
   doi = {10.1103/PhysRevLett.125.158002},
   issn = {0031-9007},
   issue = {15},
   journal = {Physical Review Letters},
   pages = {158002},
   publisher = {American Physical Society},
   title = {Flow-Driven Branching in a Frangible Porous Medium},
   volume = {125},
   year = {2020},
}

@article{Maionchi2008,
   author = {D. O. Maionchi and A. F. Morais and R. N. {Costa Filho} and J. S. {Andrade Jr.} and H. J. Herrmann},
   doi = {10.1103/PHYSREVE.77.061402/FIGURES/8/MEDIUM},
   issn = {1539-3755},
   issue = {6},
   journal = {Physical Review E},
   pages = {061402},
   publisher = {American Physical Society (APS)},
   title = {Model for erosion-deposition patterns},
   volume = {77},
   year = {2008},
}

@article{Kudrolli2016,
   author = {Arshad Kudrolli and Xavier Clotet},
   doi = {10.1103/PhysRevLett.117.028001},
   issn = {0031-9007},
   issue = {2},
   journal = {Physical Review Letters},
   pages = {028001},
   publisher = {American Physical Society},
   title = {Evolution of Porosity and Channelization of an Erosive Medium Driven by Fluid Flow},
   volume = {117},
   year = {2016},
}

@article{Graczyk2020,
   author = {Krzysztof M. Graczyk and Maciej Matyka},
   doi = {10.1038/s41598-020-78415-x},
   isbn = {0123456789},
   issn = {2045-2322},
   issue = {1},
   journal = {Scientific Reports 2020 10:1},
   pages = {1-11},
   pmid = {33293546},
   publisher = {Nature Publishing Group},
   title = {Predicting porosity, permeability, and tortuosity of porous media from images by deep learning},
   volume = {10},
   year = {2020},
}

@article{Ladd2021,
   author = {Anthony J.C. Ladd and Piotr Szymczak},
   doi = {10.1146/annurev-chembioeng-092920-102703},
   issn = {1947-5438},
   issue = {1},
   journal = {Annual Review of Chemical and Biomolecular Engineering},
   pages = {543-571},
   pmid = {33784175},
   publisher = { Annual Reviews },
   title = {Reactive Flows in Porous Media: Challenges in Theoretical and Numerical Methods},
   volume = {12},
   year = {2021},
}

@article{Oost2007,
   author = {K. {Van Oost} and T. A. Quine and G. Govers and S. {De Gryze} and J. Six and J. W. Harden and J. C. Ritchie and G. W. McCarty and G. Heckrath and C. Kosmas and J. V. Giraldez and J. R. {Marques da Silva} and R. Merckx},
   doi = {10.1126/science.1145724},
   issn = {0036-8075},
   issue = {5850},
   journal = {Science},
   pages = {626-629},
   publisher = {American Association for the Advancement of Science},
   title = {The Impact of Agricultural Soil Erosion on the Global Carbon Cycle},
   volume = {318},
   year = {2007},
}

@article{Borrelli2017,
   author = {Pasquale Borrelli and David A. Robinson and Larissa R. Fleischer and Emanuele Lugato and Cristiano Ballabio and Christine Alewell and Katrin Meusburger and Sirio Modugno and Brigitta Schütt and Vito Ferro and Vincenzo Bagarello and Kristof Van Oost and Luca Montanarella and Panos Panagos},
   doi = {10.1038/s41467-017-02142-7},
   issn = {2041-1723},
   issue = {1},
   journal = {Nature Communications},
   pages = {2013},
   pmid = {29222506},
   publisher = {Nature Publishing Group},
   title = {An assessment of the global impact of 21st century land use change on soil erosion},
   volume = {8},
   year = {2017},
}

@book{Acheson1990,
   author = {D. J. Acheson},
   isbn = {9780191059391},
   publisher = {Clarendon Press},
   title = {Elementary Fluid Dynamics},
   year = {1990},
}

@article{Carman1997,
   author = {P.C. Carman},
   title = {Fluid flow through granular beds},
   journal = {Chemical Engineering Research and Design},
   volume = {75},
   pages = {S32-S48},
   year = {1997},
   issn = {0263-8762},
   doi = {https://doi.org/10.1016/S0263-8762(97)80003-2}
}

@article{Kozeny1927,
  author={Kozeny, Josef},
  journal={Mathematisch Naturwissenschaftliche Abteilung},
  volume={136},
  pages={271--306},
  year={1927}
}

@article{Mahadevan2012,
   author = {A. Mahadevan and A. V. Orpe and Arshad Kudrolli and L. Mahadevan},
   doi = {10.1209/0295-5075/98/58003},
   issn = {0295-5075},
   issue = {5},
   journal = {EPL (Europhysics Letters)},
   month = {6},
   pages = {58003},
   publisher = {IOP Publishing},
   title = {Flow-induced channelization in a porous medium},
   volume = {98},
   year = {2012},
}

@article{Ocko2015,
   author = {Samuel A. Ocko and L. Mahadevan},
   doi = {10.1103/PhysRevLett.114.134501},
   issn = {0031-9007},
   issue = {13},
   journal = {Physical Review Letters},
   pages = {134501},
   publisher = {American Physical Society},
   title = {Feedback-Induced Phase Transitions in Active Heterogeneous Conductors},
   volume = {114},
   year = {2015},
}

@article{Mo2021,
   author = {Chaojie Mo and Richard Johnston and Luciano Navarini and Marco Ellero},
   doi = {10.1063/5.0059707},
   issn = {1070-6631},
   issue = {9},
   journal = {Physics of Fluids},
   pages = {093101},
   publisher = {AIP Publishing LLCAIP Publishing},
   title = {Modeling the effect of flow-induced mechanical erosion during coffee filtration},
   volume = {33},
   year = {2021},
}

@article{Moore2023,
   author = {Nicholas J. Moore and Jake Cherry and Shang-Huan Chiu and Bryan D. Quaife},
   doi = {10.1016/j.physd.2022.133634},
   issn = {01672789},
   journal = {Physica D: Nonlinear Phenomena},
   pages = {133634},
   title = {How fluid-mechanical erosion creates anisotropic porous media},
   volume = {445},
   year = {2023},
}

@article{Mondal2019,
   author = {Raka Mondal and Sourav Mondal and Krishnasri V. Kurada and Saikat Bhattacharjee and Sourav Sengupta and Mrinmoy Mondal and Sankha Karmakar and Sirshendu De and Ian M. Griffiths},
   doi = {10.1016/J.CES.2019.115205},
   issn = {0009-2509},
   journal = {Chemical Engineering Science},
   pages = {115205},
   publisher = {Pergamon},
   title = {Modelling the transport and adsorption dynamics of arsenic in a soil bed filter},
   volume = {210},
   year = {2019},
}

@article{Dalwadi2016,
   author = {M. P. Dalwadi and M. Bruna and I. M. Griffiths},
   doi = {10.1017/jfm.2016.656},
   issn = {0022-1120},
   journal = {Journal of Fluid Mechanics},
   pages = {264-289},
   publisher = {Cambridge University Press},
   title = {A multiscale method to calculate filter blockage},
   volume = {809},
   year = {2016},
}

@article{Banavar1997,
   author = {Jayanth R. Banavar and Francesca Colaiori and Allesandro Flammini and Achille Giacometti and Amos Maritan and Andrea Rinaldo},
   doi = {10.1103/PhysRevLett.78.4522},
   issn = {10797114},
   issue = {23},
   journal = {Physical Review Letters},
   pages = {4522},
   publisher = {American Physical Society},
   title = {Sculpting of a Fractal River Basin},
   volume = {78},
   year = {1997},
}

@article{Lefebvre2016,
   author = {Gautier Lefebvre and Aymeric Merceron and Pierre Jop},
   doi = {10.1103/PhysRevLett.116.068002},
   issn = {0031-9007},
   issue = {6},
   journal = {Physical Review Letters},
   pages = {068002},
   publisher = {American Physical Society},
   title = {Interfacial Instability during Granular Erosion},
   volume = {116},
   year = {2016},
}

@article{Ferlito2006,
   author = {Carmelo Ferlito and Jens Siewert},
   doi = {10.1103/PhysRevLett.96.028501},
   issn = {0031-9007},
   issue = {2},
   journal = {Physical Review Letters},
   pages = {028501},
   publisher = {American Physical Society},
   title = {Lava Channel Formation during the 2001 Eruption on Mount Etna: Evidence for Mechanical Erosion},
   volume = {96},
   year = {2006},
}

@article{Zwietering1954,
   author = {P. Zwietering and H. L. T. Koks},
   doi = {10.1038/173683a0},
   issn = {0028-0836},
   issue = {4406},
   journal = {Nature},
   pages = {683-684},
   publisher = {Nature Publishing Group},
   title = {Pore-Size Distribution of Porous Iron},
   volume = {173},
   year = {1954},
}

@article{Zhu2019,
   author = {Jie Zhu and Rui Zhang and Yang Zhang and Fa He},
   doi = {10.1038/s41598-019-53828-5},
   issn = {2045-2322},
   issue = {1},
   journal = {Scientific Reports},
   pages = {17191},
   pmid = {31748617},
   publisher = {Nature Publishing Group},
   title = {The fractal characteristics of pore size distribution in cement-based materials and its effect on gas permeability},
   volume = {9},
   year = {2019},
}

@article{Sun2020,
   author = {Bin Sun and Qing Yang and Jie Zhu and Tangsha Shao and Yuhang Yang and Chenyu Hou and Guiyou Li},
   doi = {10.1038/s41598-020-79338-3},
   isbn = {0123456789},
   issn = {2045-2322},
   issue = {1},
   journal = {Scientific Reports},
   pages = {22353},
   pmid = {33339868},
   publisher = {Nature Publishing Group},
   title = {Pore size distributions and pore multifractal characteristics of medium and low-rank coals},
   volume = {10},
   year = {2020},
}

@article{Hilton2020,
   author = {Robert G. Hilton and A. Joshua West},
   doi = {10.1038/s43017-020-0058-6},
   issn = {2662-138X},
   issue = {6},
   journal = {Nature Reviews Earth \& Environment 2020 1:6},
   pages = {284-299},
   publisher = {Nature Publishing Group},
   title = {Mountains, erosion and the carbon cycle},
   volume = {1},
   year = {2020},
}

@article{Andrade2009,
   author = {J. S. {Andrade Jr.} and E. A. Oliveira and A. A. Moreira and H. J. Herrmann},
   doi = {10.1103/PhysRevLett.103.225503},
   issn = {0031-9007},
   issue = {22},
   journal = {Physical Review Letters},
   pages = {225503},
   publisher = {American Physical Society},
   title = {Fracturing the Optimal Paths},
   volume = {103},
   year = {2009},
}

@article{Maliva2015,
   author = {Robert G. Maliva and Rolf Herrmann and Kapo Coulibaly and Weixing Guo},
   doi = {10.1007/s12665-014-3167-z},
   issn = {1866-6280},
   issue = {12},
   journal = {Environmental Earth Sciences},
   pages = {7759-7767},
   publisher = {Springer Verlag},
   title = {Advanced aquifer characterization for optimization of managed aquifer recharge},
   volume = {73},
   year = {2015},
}

@article{Goretta1999,
   author = {K.C Goretta and M.L Burdt and M.M Cuber and L.A Perry and D Singh and A.S Wagh and J.L Routbort and W.J Weber},
   doi = {10.1016/S0043-1648(98)00339-1},
   issn = {00431648},
   issue = {1},
   journal = {Wear},
   pages = {106-112},
   publisher = {Elsevier},
   title = {Solid-particle erosion of Portland cement and concrete},
   volume = {224},
   year = {1999},
}

@article{Birien2023,
   author = {Tom Birien and Francis Gauthier},
   doi = {10.1016/j.geomorph.2022.108518},
   issn = {0169555X},
   journal = {Geomorphology},
   pages = {108518},
   publisher = {Elsevier},
   title = {Influence of climate-dependent variables on deformation and differential erosion of stratified sedimentary rocks},
   volume = {421},
   year = {2023},
}

@book{Sahimi2011,
   author = {Muhammad Sahimi},
   city = {Weinheim, Germany},
   doi = {10.1002/9783527636693},
   edition = {2},
   isbn = {9783527404858},
   publisher = {Wiley-VCH},
   title = {Flow and Transport in Porous Media and Fractured Rock: From Classical Methods to Modern Approaches},
   year = {2011},
}

@article{Sahimi1993,
   author = {Muhammad Sahimi},
   doi = {10.1103/RevModPhys.65.1393},
   issn = {00346861},
   issue = {4},
   journal = {Reviews of Modern Physics},
   pages = {1393-1534},
   publisher = {American Physical Society},
   title = {Flow phenomena in rocks: From continuum models to fractals, percolation, cellular automata, and simulated annealing},
   volume = {65},
   year = {1993},
}

@article{Rassamdana1996,
   author = {Hossein Rassamdana and Bahram Dabir and Mehdi Nematy and Minoo Farhani and Muhammad Sahimi},
   doi = {10.1002/aic.690420104},
   issn = {0001-1541},
   issue = {1},
   journal = {AIChE Journal},
   pages = {10-22},
   publisher = {John Wiley & Sons, Ltd},
   title = {Asphalt flocculation and deposition: I. The onset of precipitation},
   volume = {42},
   year = {1996},
}

@article{Zhang2023,
   author = {Dawang Zhang and James M. Campbell and Jon A. Eriksen and Eirik G. Flekkøy and Knut Jørgen Måløy and Christopher W. MacMinn and Bjørnar Sandnes},
   doi = {10.1038/s41467-023-38648-6},
   issn = {2041-1723},
   issue = {1},
   journal = {Nature Communications},
   pages = {3044},
   pmid = {37236971},
   publisher = {Nature Publishing Group},
   title = {Frictional fluid instabilities shaped by viscous forces},
   volume = {14},
   year = {2023},
}

@article{Xu2018,
   author = {Le Xu and Benjy Marks and Renaud Toussaint and Eirik G. Flekkøy and Knut J. Måløy},
   doi = {10.3389/FPHY.2018.00029/BIBTEX},
   issn = {2296424X},
   issue = {APR},
   journal = {Frontiers in Physics},
   pages = {356235},
   publisher = {Frontiers Media S.A.},
   title = {Dispersion in fractures with ramified dissolution patterns},
   volume = {6},
   year = {2018},
}

@article{Soares1999,
   author = {José S. {Andrade Jr.} and U.M.S. Costa and H.A. Makse and H.E. Stanley},
   doi = {10.1016/S0378-4371(98)00624-4},
   issn = {03784371},
   issue = {1-4},
   journal = {Physica A: Statistical Mechanics and its Applications},
   pages = {420-424},
   publisher = {American Physical Society},
   title = {The role of inertia on fluid flow through disordered porous media},
   volume = {266},
   year = {1999},
}

@article{Cerasi1998,
   author = {P. Cerasi and P. Mills},
   doi = {10.1103/PhysRevE.58.6051},
   issn = {1063651X},
   issue = {5},
   journal = {Physical Review E},
   pages = {6051},
   publisher = {American Physical Society},
   title = {Insights in erosion instabilities in nonconsolidated porous media},
   volume = {58},
   year = {1998}
}

@article{Kahza2024,
   author = {Hamad El Kahza and Pejman Sanaei},
   doi = {10.1103/PHYSREVFLUIDS.9.024301/DELIVERABLE/66E52895-1D8B-4289-8507-560834530049},
   issn = {2469990X},
   issue = {2},
   journal = {Physical Review Fluids},
   pages = {024301},
   publisher = {American Physical Society},
   title = {Mathematical modeling of erosion and deposition in porous media},
   volume = {9},
   year = {2024}
}

@article{Tauber2026,
   author = {J. Tauber and J. Asnacios and L. Mahadevan},
   doi = {10.1103/lps4-8j6d},
   issn = {2469-990X},
   issue = {2},
   journal = {Physical Review Fluids},
   pages = {023301},
   title = {Self-organized breakthrough morphodynamics in fluid-driven branching},
   volume = {11},
   year = {2026}
}

@article{Araujo2006,
  title = {Critical Role of Gravity in Filters},
  author = {Ara\'ujo, A. D. and {Andrade Jr.}, J. S. and Herrmann, H. J.},
  journal = {Phys. Rev. Lett.},
  volume = {97},
  issue = {13},
  pages = {138001},
  year = {2006},
  publisher = {American Physical Society},
  doi = {10.1103/PhysRevLett.97.138001}
}

@article{Andrade1999,
  title = {Inertial Effects on Fluid Flow through Disordered Porous Media},
  author = {{Andrade Jr.}, J. S. and Costa, U. M. S. and Almeida, M. P. and Makse, H. A. and Stanley, H. E.},
  journal = {Phys. Rev. Lett.},
  volume = {82},
  issue = {26},
  pages = {5249--5252},
  year = {1999},
  publisher = {American Physical Society},
  doi = {10.1103/PhysRevLett.82.5249}
}

@article{Lee2025,
   author = {Sang Hyun Lee and Marcel Moura and Shreya Srivastava and Cara Santelli and Peter K. Kang},
   doi = {10.1038/s41567-025-03020-6},
   issn = {1745-2481},
   issue = {11},
   journal = {Nature Physics 2025 21:11},
   pages = {1719-1727},
   publisher = {Nature Publishing Group},
   title = {Filamentous fungi control multiphase flow and fluid distribution in porous media},
   volume = {21},
   year = {2025}
}

@Misc{SM,
   note = {See Supplemental Material at ... for supplemental figures, and parameter table.},
   url = {...},
}
\bibliographystyle{apsrev4-2}

\section{End Matter}

\subsection{Wall shear stress at the pore scale}
\label{sup:wss}

\begin{figure}
	\centering
	\includegraphics[width=\linewidth]{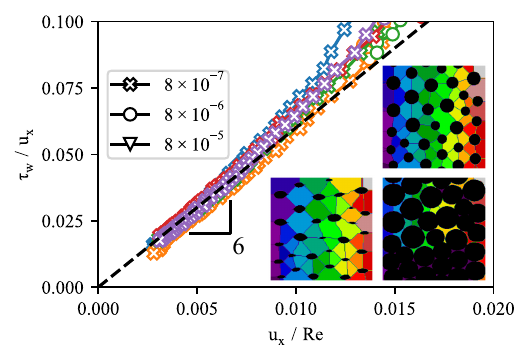}
	\caption[Pore-scale wall shear stress as a function of the average fluid velocity]{\textbf{Pore-scale wall shear stress as a function of the average fluid velocity.} The symbols correspond to different erosion/deposition thresholds and the colors to different samples. The dashed line is given by the analytical prediction, Eq.~\eqref{eq:wss}. The inset contains examples of a porous medium with the particles (black) and the water basins (colors) determined using the watershed method for the initial system (top), after erosion (bottom left), and after deposition (bottom right). The lines separating the basins correspond to the pore size between each pair of particles used to determine the Reynolds number ($\textrm{Re}$).}
	\label{fig:wss_validation}
\end{figure}

The Poiseuille flow past two parallel plates separated by a distance $h$ given by
\begin{equation}
	u(y)=\frac{G}{2 \rho \nu} y (h - y).
\end{equation}
Thus, assuming a constant flow rate, the wall shear stress can be written as,
\begin{equation}
	\tau_w = \rho \nu \left.\frac{\partial u}{\partial y}\right|_{y=0} = 6 \rho \nu \frac{\langle u\rangle}{h} = 6 \rho \frac{\langle u\rangle^2}{\textrm{Re}},
\end{equation}
where $\langle u\rangle$ is the volume average fluid velocity. Assuming the capillary model, this expression should be valid for a more complex medium. In this case the Reynolds number is composed by the average fluid velocity and the average pore size. The average pore size was determined using the watershed method~\cite{Rabbani2014}. In the insets of Fig.~\ref{fig:wss_validation} are the result of the watershed for the initial system (top), after erosion (bottom left) and after deposition (bottom right). The average pore size corresponds to the average distance between particles, marked with lines limiting the different basins (colors). The wall shear stress in pore-scale simulations at low Reynolds flow, $\textrm{Re}\ll 1$, across packed spheres during erosion and deposition agrees with this approximation. Thus, the rescaled wall shear stress increases linearly with the ratio between the velocity and Reynolds number, see Fig.~\ref{fig:wss_validation}. We simulated five different samples (colors). For $\tau = 8\times10^{-5}$ l.u. deposition dominates, for $\tau = 8\times10^{-7}$ l.u. erosion dominates, and for $\tau = 8\times10^{-6}$ l.u. erosion and deposition compete and there is the formation of channels. For all thresholds the lines collapse and closely follow the theoretical prediction (dashed line). After deposition, some pores are blocked and unconnected regions appear, see the bottom right inset of Fig.~\ref{fig:wss_validation}, which impacts the accuracy of the estimation of average pore size, and so the accuracy of the model. Overall, these simulations show that the capillary approximation is valid to determine the wall shear stress for the case of a bed of packed circles.

\subsubsection{Simulation of the fluid flow through circles}

To simulate the fluid flow we used the lattice Boltzmann method~\cite{Kruger2017} with the MRT collision operator and a ghost method for the boundaries between particles and fluid~\cite{Mei1999}. A body force of $10^{-6}$ l.u. is imposed on all fluid nodes. The fluid viscosity was set such that $Re \ll 1$. The boundaries evolve according to Eq.~\eqref{eq:cap_evolution} and $\kappa = 10$. The erosion and deposition thresholds are equal and were varied to control whether erosion or deposition dominates the dynamics. The domain size is $256 \times 384$ nodes and the first and last quarter have no particles. The particles radii follow a Gaussian with mean $20$ nodes and $5\%$ dispersion. The initial position of the particles is determined by compressing the bed until a size of $256^2$ is reached. After compression the particles radii are rescaled by $75\%$ to allow for the fluid flow.

\subsection{Wall shear stress for 3D systems}

The Poiseuille flow through a cylinder of radius $a$ is
\begin{equation}
	u(r)=\frac{G}{4 \rho \nu} (a^2 - r^2),
\end{equation}
thus, the wall shear stress, assuming constant flow rate, is
\begin{equation}
	\tau_w = 4 \rho \nu \frac{\langle u\rangle}{R} = 4 \rho \frac{\langle u\rangle^2}{\textrm{Re}}.
\end{equation}
Given the extensive validations of the capillary model, and our validation of the wall shear stress approximation for 2D geometries, we expect this equation to be valid for compacted spheres. To implement the erosion/deposition model, described on the main text, the porosity evolution needs to be corrected. For 3D geometries the capillaries are cylinders, and thus, the porosity changes with the capillary radius according to
\begin{equation}
	\frac{\dot{\phi}}{\phi}=\frac{2\dot{a}}{a} .
\end{equation}

\subsection{Erosion resistance and porosity distribution}

The distribution of erosion resistance $\tau$ follows hyperbolic disorder, Eq.~\eqref{eq:power_law}. Numerically, the $\tau$ values are generated using the transformation method, $\tau_i/\tau_{\min} = \exp\left[-\beta_\tau\left(x_i - 1\right)\right]$, where $x_i$ is a random number uniformly distributed in $[0,1[$, and $\beta_\tau$ is the disorder parameter. For the case of disorder in the porosity, we follow the same methodology but generate the porosity values with $\phi_i/\phi_{\max} = \exp\left[+\beta_\phi\left(x_i - 1\right)\right]$. In all simulations, we avoid discontinuities by bounding the porosity to the interval $\phi \in [0.1, 0.9]$.

\clearpage
\onecolumngrid

\renewcommand{\thepage}{S\arabic{page}}
\setcounter{page}{1}
\renewcommand{\thesection}{S\arabic{section}}
\setcounter{section}{0}
\renewcommand{\thetable}{S\arabic{table}}
\renewcommand{\thefigure}{S\arabic{figure}}
\setcounter{figure}{0}
\renewcommand{\theequation}{S\arabic{equation}}
\setcounter{equation}{0}
\def\lp{\left(}
\def\rp{\right)}
\def\lb{\left[}
\def\rb{\right]}

\section*{Supplementary Material}

\begin{figure}[!ht]
	\centering
	\includegraphics[width=\linewidth]{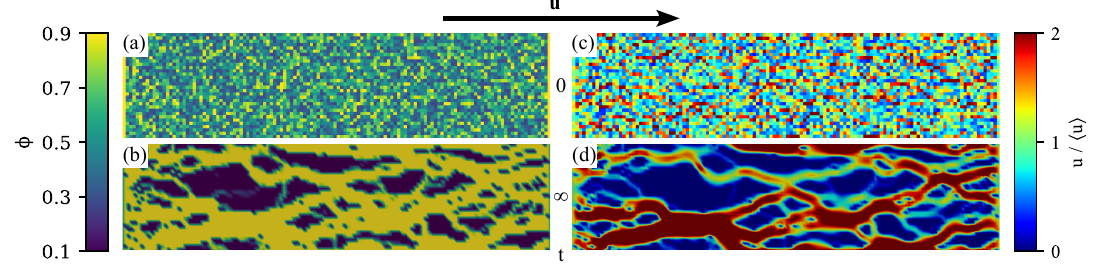}
    \caption{(a)-(b) Porosity and (c)-(d) velocity field for the initial time (top) and steady-state (bottom) for $\beta_\phi=1$. The velocity field strongly depends on the porosity field.}
	\label{fig:supple_porosity}
\end{figure}

\begin{figure}[!ht]
	\centering
	\includegraphics[width=\linewidth]{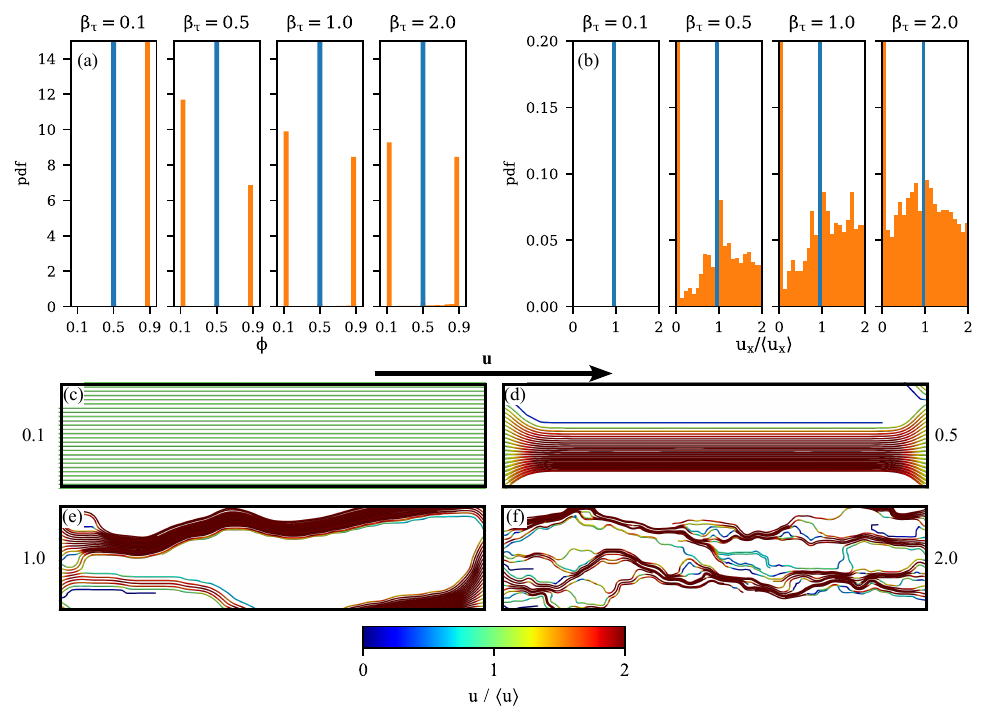}
	\caption{Histogram of the porosity (a) and velocity (b) for the initial time (blue) and steady-state (orange) for different values of $\beta_\tau$. The erosion and deposition lead the porosity to a bimodal distribution of nodes with $\phi=0.1$ and $0.9$. The velocity distribution gets wider as some nodes have high porosity and thus concentrate most of the fluid flow. (c)-(f) Velocity streamlines for different $\beta_\tau$, the color represents the velocity magnitude. For weak disorder the flow is homogeneous (c). Increasing the disorder results in the channelization of the flow (d). For strong disorder the flow occurs across several channels (e), beyond this point increasing the disorder increases the number of channels (f).}
	\label{fig:supple_threshold}
\end{figure}

\begin{figure}[!ht]
	\centering
	\includegraphics[width=.5\linewidth]{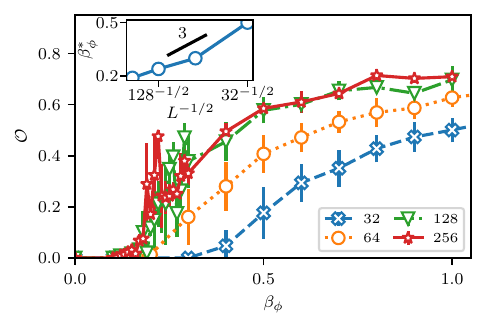}
	\caption{Disorder on the initial porosity for $\tau=6\times 10^{-8}$ l.u. Channelization parameter $\mathcal{O}$ as a function of the strength of disorder in the initial porosity, for different system sizes (colors and markers). Each point is an average over $N$ samples, that decrease with the system size with a minimum of $N=4$, in the steady state. Error bars correspond to the standard error. Unlike the case of disorder in the erosion resistance, very weak disorder is enough to induce channelization. The inset contains the log-scale plot of the strength of disorder that corresponds to the maximum second momentum of $\mathcal{O}$ as a function of the inverse square root of the system size $L^{-1/2}$.}
	\label{fig:supple_lower_tau_cri}
\end{figure}

\twocolumngrid

\begin{table}[!ht]
	\label{tab:parameters}
	\caption{List of simulation parameters. All parameters are in lattice units, a set of units that can be converted into real ones using dimensionless numbers, such as the Reynolds number.}
	\begin{ruledtabular}
		\begin{tabular}{lcr}
			Parameter                    &            Symbol            &     Value \\
			\colrule
			Imposed velocity & $\langle \mathbf{u} \rangle$ & $10^{-6}$ \\
			Erosion/deposition threshold &            $\tau$            &       $10^{-7}$ \\
			Erosion/deposition rate      &           $\kappa$           & $1$ \\
			Fluid density                &            $\rho$            &             $1$ \\
			Kinematic viscosity          &            $\nu$             &           $0.1$ \\
			Initial capillary radius     &            $a_i$             &              1
		\end{tabular}
	\end{ruledtabular}
\end{table}

\end{document}